\documentclass[conference]{IEEEtran}
\ifCLASSINFOpdf
\else
\fi

\usepackage{graphics}
\usepackage[pdftex]{graphicx}
\usepackage{subfigure}
\usepackage{multirow}
\usepackage{textcomp}
\usepackage{array}
\usepackage{url}
\makeatletter
\g@addto@macro{\UrlBreaks}{\UrlOrds}
\makeatother

\begin{document}
%
\title{Know Your Master: Driver Profiling-based Anti-theft Method}

\author{\IEEEauthorblockN{Byung Il Kwak}
\IEEEauthorblockA{Korea University\\
Seoul, Republic of Korea\\
Email: kwacka12@korea.ac.kr}
\and
\IEEEauthorblockN{JiYoung Woo}
\IEEEauthorblockA{Soonchunhyang University\\
Asan, Republic of Korea\\
Email: jywoo@sch.ac.kr}
\and
\IEEEauthorblockN{Huy Kang Kim\\ }
\IEEEauthorblockA{Korea University\\
Seoul, Republic of Korea\\
Email: cenda@korea.ac.kr}}


%


\maketitle

\begin{abstract}
Although many anti-theft technologies are implemented, auto-theft is still increasing. Also, security vulnerabilities of cars can be used for auto-theft by neutralizing anti-theft system. This keyless auto-theft attack will be increased as cars adopt computerized electronic devices more. 
To detect auto-theft efficiently, we propose the driver verification method that analyzes driving patterns using measurements from the sensor in the vehicle. In our model, we add mechanical features of automotive parts that are excluded in previous works, but can be differentiated by drivers' driving behaviors.
We design the model that uses significant features through feature selection to reduce the time cost of feature processing and improve the detection performance.
Further, we enrich the feature set by deriving statistical features such as mean, median, and standard deviation. This minimizes the effect of fluctuation of feature values per driver and finally generates the reliable model. We also analyze the effect of the size of sliding window on performance to detect the time point when the detection becomes reliable and to inform owners the theft event as soon as possible. 

We apply our model with real driving and show the contribution of our work to the literature of driver identification. 
\end{abstract}


%
\IEEEpeerreviewmaketitle

\section{Introduction}

The number of stolen vehicles is increasing every day. FBI Uniform Crime Reports said that 689,527 vehicles were stolen in the United States in 2014. This means that auto-theft occurred every 46 seconds in 2014.
Cars are increasingly exposed to theft. Cars today have many computers on board. Further, as ICT has been developing, cars become increasingly connected to the Internet. 
This trend will be associated with serious risks as follows.
The software operating cars are likely to have many bugs and vulnerabilities. 
It is technically feasible that attackers can exploit vulnerabilities through the Internet connection. Already, many vulnerabilities are being discovered. Even though threats against modern cars are severe, the security system is hardly installed in cars. 
Vulnerabilities of connected cars will increase auto-theft that is one of the threats. 
Top-of-the-range vehicles are targeted by thieves who simply drive off after bypassing security devices by hacking on-board computers \cite{add1}. Connected cars contain computer components, so keyless techniques are used to steal cars. One of the current techniques is breaking into the vehicle and plugging a laptop into the hidden diagnostic socket \cite{add1}.
The other exploit method is that attackers can simply get owners to install malware into their smartphones working as a door lock and make the door open. BMW recently patched its ConnectedDrive system as researchers showed that it was possible to get wireless access to the air conditioning and door lock of cars \cite{add2}.
The threats being discovered will be realized and the security of connected cars will become more important as more cars are connected to the Internet. Gartner reports that there will be a quarter of a billion connected vehicles by 2020 \cite{refer0}.

As the cyber-attacks become diverse and intelligent, the protection techniques are not enough to response to them. To detect novel attacks, the data-mining techniques are widely used. Data-mining techniques are also useful to detect car-theft attempts because drivers have own habits of driving, so they exhibit unique driving patterns. To detect vehicle theft regardless of attack methods, we propose an authentication method based on driver behaviors that are difficult to detour. 

Previous works proposed authentication methods focusing biometrics such as password, fingerprinting, iris recognition, and face recognition. These methods are hard to have high accuracy, and they fail to be applied for other security issues in the connected car. 
The authentication based on behavioral characteristics would be an alternative to these biometric-based authentication methods.
It is difficult to bypass the various driver's driving characteristics such as vehicle interval, sudden unintended acceleration, maximum driving speed, the number of brakes, and the angle of steering handle. Thus, analyzing driver's running pattern is a good way to authenticate the driver.
Previous works collected driving data from cars and used it in driver profiling \cite{refer9}. However, these studies have room for improvement in enriching features and analysis dimensions such as temporal perspective. Previous works focused on the detection accuracy, but ignored the timeliness.
In this work, we develop the driver verification model to achieve highly accurate automatic detection of auto-theft including other unauthorized usages. In addition, we examine the temporal data and train the model with different analysis periods. This will inform us how long data should be collected and when the detector should inform owners break-in of unauthorized users.

Our contributions are as follows:
\begin{enumerate}[\IEEEsetlabelwidth{3)}]
\item We analyze drivers' real driving data. Moreover, we collect the data from three road types of the motor way, city way and parking lot. The driving data were repeatedly collected by ten drivers.
\item We classify drivers based on behavior characteristics that are difficult to detour. For high accuracy, we enrich the feature set encompassing behavior features related to braking and accelerating, which are widely used in the previous works, and mechanical features derived from driver's behaviors.
\item We design the model to consider the significant feature through feature selection. This reduces the time cost of feature processing and improves the detection performance.
\item We enrich the feature set by deriving statistical features such as mean, median, and standard deviation. This minimizes the effect of fluctuation of feature values per driver and finally generates a reliable model.
\item We process sliding window to detect the time point when the detection becomes reliable and to inform owners theft event as soon as possible. 

\end{enumerate}

We make a decision whether a driver is the authenticated user or not through analyzing the driving data. The second section summarizes studies of driver profiling through the data analysis. The third section provides the method of driver authentication and classification result. Moreover, the fourth section presents discussions. Finally, the fifth section provides conclusion and future work of this study.

\section{Related works}

As the response of cyber-attack in the vehicle, security technology in the vehicle has been studied and developed. There are many ongoing international projects of vehicle security. E.g., SEVECOM (SEcure VEhicle COMmunication), EVITA (E-safety Vehicle Intrusion Protected Applications), PRESERVE (Preparing Secure V2X Communication Systems), IntelliDrive. SEVECOM project defines attacks and threats in the vehicle network. For dealing with them, SEVECOM has been researching cryptography algorithm and cryptography protocols in the vehicle network. For dealing with them, SEVECOM has been researching cryptography algorithm and cryptography protocol. EVITA project built Hardware Security Module (HSM) and developed the vehicle inner network traffic encryption. EVITA project built Hardware Security Module (HSM) and developed the vehicle inner network traffic encryption. PRESERVE project integrates the projects of vehicle security in progress across Europe and constitutes V2X (Vehicle to everything) security system. V2X focuses on data sharing technology between car and vehicle, or between vehicle and traffic control infrastructure. IntelliDrive Project developed in America aims to increase safety, mobility, and eco-friendly properties through the constructing system for inter-vehicle communication. As the results of these efforts, ISO 26262 \cite{refer2} that is a Functional Safety standard for road vehicles and AUTOSAR \cite{refer3} that is automotive open system architecture have been established for the safety of vehicles.

In the past, data collection in the real world had a limitation because it was difficult to equip the sensors in vehicles. Kawaguchi et al. \cite{refer15} proposed the methods for data collection. They installed many sensors, video camera, GPS antenna, and desktop PC for data collection. Collected data include face image, vehicle velocity, and brake pedal pressure. These data can be used for various purposes such as driver behavior analysis \cite{refer16}, intrusion detection in vehicle CAN network \cite{refer17} \cite{refer18}, and anti-theft \cite{refer19}.

\begin{table*}[ht!]
\renewcommand{\arraystretch}{1.3}
\vspace{-3mm}
\caption{Research on Driver Classification}
\label{table1}
\centering
\begin{tabular}{|m{2.8cm}|m{3.4cm}|m{6.7cm}|m{3cm}|}
\hline
\multicolumn{1}{|c|}{Research}	&	\multicolumn{1}{c|}{Collected dataset}	&	\multicolumn{1}{c|}{Feature}	&	\multicolumn{1}{c|}{Classification algorithm}   \\
\hline
Wakita et al. \cite{refer10}, 2006	&	Driving simulation data	&	Vehicle speed, following distance from vehicle ahead, accelerator pedal pressure, brake pedal pressure 	&	Gaussian Mixture Model (GMM)  \\
\hline
Meng et al. \cite{refer5}, 2006	&	Driving simulation data	&	Acceleration, brake, steering wheel &	HMM  \\
\hline
Miyajima et al. \cite{refer4}, 2007	&	Driving simulation data, In-vehicle's sensor data	&	Vehicle speed, brake pedal position, gas pedal position, following distance from vehicle ahead, brake pedal pressure, gas pedal pressure, engine speed  	&	GMM  \\
\hline
Nishiwaki et al. \cite{refer7}, 2007	&	In-vehicle's sensor data 	&	Brake pedal pressure, gas pedal pressure	&	GMM  \\
\hline

Choi et al. \cite{refer6}, 2007	&	In-vehicle's CAN network data	& Steering wheel, vehicle speed, engine speed, brake position  	&	GMM, HMM  \\
\hline
Wahab et al. \cite{refer12}, 2009	&	In-vehicle's sensor data, video streams, voice streams	&	Accelerator pedal pressure, brake pedal pressure	&	MLP, statistical method, Fuzzy-neural-network (FNN)  \\
\hline

Kedar-Dongarkar et al. \cite{refer11}, 2012	&	In-vehicle's CAN network data	&	Vehicle speed, acceleration, torque, accelerator pedal pressure, steering wheel, brake pedal pressure	&	Statistical method  \\
\hline

Van Ly et al. \cite{refer13}, 2013	&	In-vehicle's CAN network data	&	Acceleration, brake, turning signal 	&	K-means, SVM  \\
\hline
Zhang et al. \cite{refer8}, 2014	&	Driving simulation data 	&	Acceleration, steering wheel	&	HMM  \\
\hline
Enev et al. \cite{refer9}, 2016	&	In-vehicle's CAN network data	&	Acceleration, brake, steering wheel, vehicle speed, engine speed, gear shift, yaw rate, shaft angular velocity, engine torque, fuel consumption, throttle position, turn signal 	&	SVM, Random Forest, Naive Bayes, KNN  \\
\hline

\end{tabular}
\vspace{-5mm}
\end{table*}
 
Table \ref{table1} lists the studies related to driver profiling and driver classification based on data analysis. Wakita et al. \cite{refer10}, Miyajima et al. \cite{refer4} and Y. Nishiwaki et al. \cite{refer7} proposed driver identification methods based on driving behavior signals. Wakita et al. \cite{refer10} generated the simulation data and compared the identification performance using different models. Miyajima et al. \cite{refer4} obtained simulation data from vehicle sensors and modeled driving behaviors as car-following and pedal operation patterns. Y. Nishiwaki et al. \cite{refer7} used vehicle sensor data and modeled the distribution of cepstral coefficients of brake and gas pedal signals.

Meng et al. \cite{refer5} studied driver's driving pattern in a game simulation. This study used HMM for modeling of drivers' dynamic behaviors using the features including acceleration, brake, and steering wheel data. HMM recognizes a pattern of data and infers the probability of a particular output sequence using the sequential data. 

HMM outperforms other classification algorithms, but the performance of HMM largely depends on the temporal data. This research also has a limitation that it performed experiments in an environment of computer simulation. 

Choi et al. \cite{refer6} researched driver actions for driver distraction detection and driver identification using both GMM and HMM frameworks on CAN network traffics of vehicles. What the strength of this works is that it performed modeling using driver's behaviors from driving using CAN that is a communication bus for transporting small control messages in real-time.

Wahab et al. \cite{refer12} performed modeling of individual driving behaviors and identified features that are efficient and effective to profile each driver. They conducted the feature extraction based on Gaussian mixture model and wavelet transform and showed that accelerator and brake pedal pressure are effective and efficient for profiling. Moreover, the research compared Multilayer Perceptron (MLP), FNN, and statistical method in detection performance.

Kedar-Dongarkar et al. \cite{refer11} applied driver classification in optimizing energy usage in a vehicle. The authors categorized drivers into three types of aggressive, moderate and conservative according to driving patterns. The features used in this work encompass vehicle speed, acceleration, torque, accelerator pedal, steering wheel angle and brake pedal pressure.
 
Van Ly et al. \cite{refer13} pointed out that driver classification can be potentially used as inertial sensors between drivers. This research used the simulation data collected by In-vehicle CAN network. The authors compared braking and turning actions, derived features from acceleration, brake and turning signals and used K-means and SVM algorithms. 

Zhang et al. \cite{refer8} researched individual characteristic of driver's running behavior. Moreover, the research built the individual models for the accuracy of driving behavior model. The research used HMM algorithm and included many features encompassing throttle, steering wheel, accelerator, brake, clutch, and gear. The research aimed to analyze the characteristics of individual drivers.

Enev et al. \cite{refer9} classified drivers through the driving data collected on the real road. This research divided the range into two driving sections of running section and parking section. They derived brake pedal position, steering wheel angle, longitudinal acceleration, turning speed, driving speed, current gear, acceleration pedal position, engine speed, maximum engine torque, fuel consumption rate, and throttle position. This work has a contribution in enriching features derived from driving and parking. This study used SVM, Random Forest, Naive Bayes, and k-Nearest Neighbor (KNN) algorithms.

\begin{figure}[b!]
\centering
\vspace{-7mm}
\includegraphics[width=0.9\columnwidth]{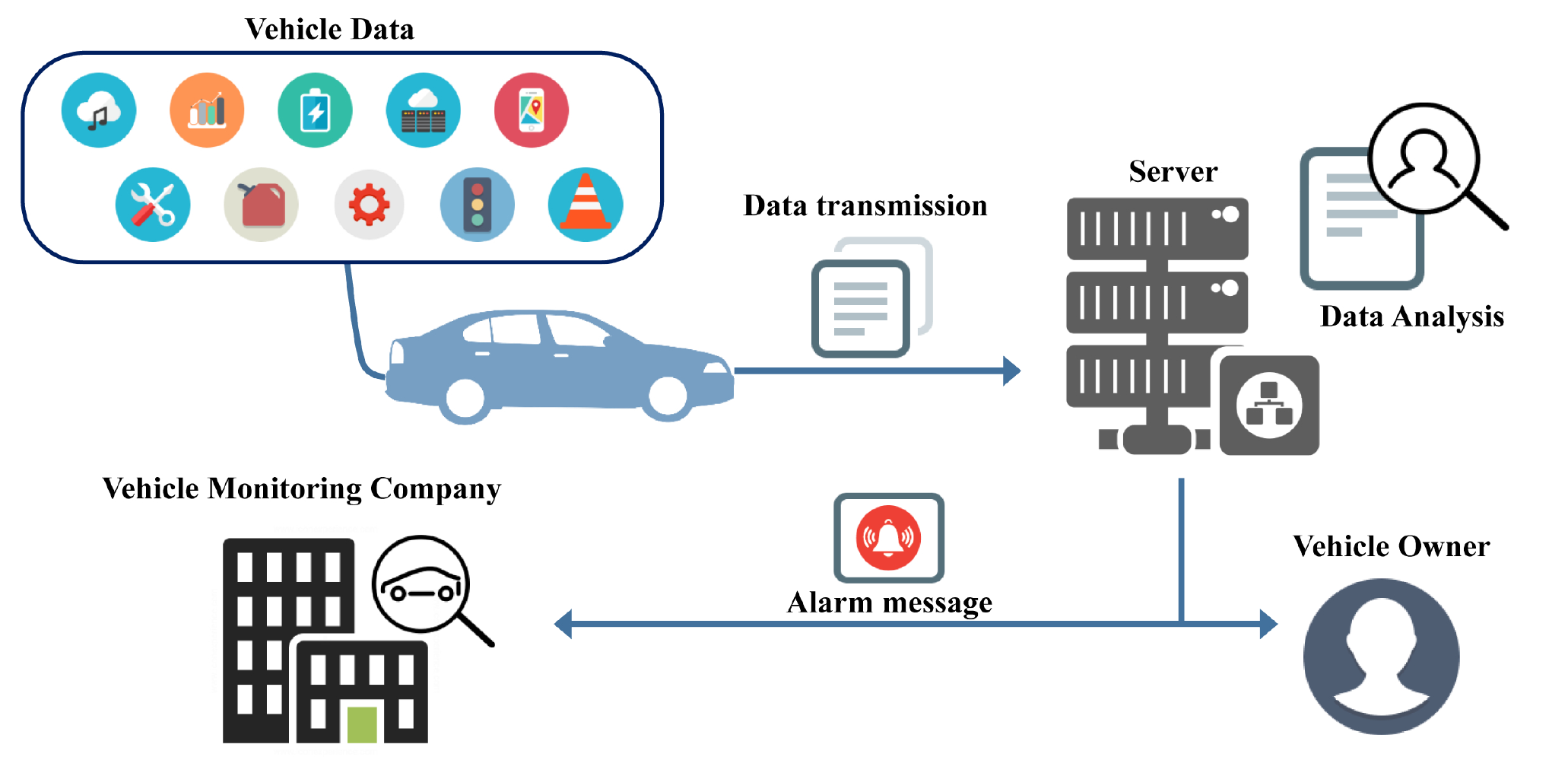}
\vspace{-3mm}
\caption{Security as a Vehicle Service in Anti-Theft}
\label{figure0}
\vspace{-2mm}
\end{figure}

\begin{figure*}[ht!]
\centering
\vspace{-3mm}
\includegraphics[width=\textwidth]{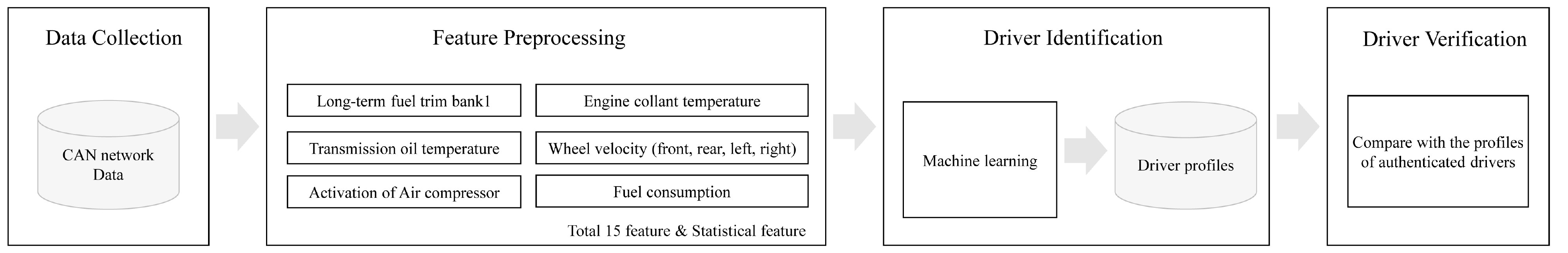}
\vspace{-5mm}
\caption{Driver Verification Framework}
\label{figure1}
\vspace{-5mm}
\end{figure*}

From thoroughly reviewing related works, authentication based on driver's driving pattern analysis is found to be an efficient method of driver identification and recognition. It can be further used in protecting privacy in connected cars. Most previous studies focused on the features that cognitively characterize driving behaviors. The features adopted in most previous works are directly extracted from the accelerator, brake, and steering handle. Previous works performed complex preprocessing, for example, the method using HMM, on these features to improve performance. 
In this work, we aim to enrich the feature set by thorough examination of CAN data for a rather easy feature extraction. 
We will also consider the mechanical features from automotive parts such as engine, fuel, and transmission. Also, we will construct real driving data from several driver's repetitive driving. 

\section{Attack Model}

We suppose the situation that the car has the computers on board and these computers are connected to the internet. The car has many embedded systems and controller units (ECUs), which communicate using the CAN (Controller Area Network) protocol. 
Cars today have infotainment console that is connected to CANBus Network and is also connected to the external through the channels like USB, Bluetooth, cellular, Wi-Fi and so on. Once attackers get access to the internal system, they can manipulate the communication within a car based on their knowledge built by prior-reversing on CAN packets for the unlocking door. 
Thus, attackers access the vehicle and drive without car owner's key. 
Moreover, they are able to bypass the alarm system in vehicles. 
If they target the vehicle, it takes a little time to steal cars.

We propose a concept of security as a vehicle service for anti-theft perspective as shown in Fig. \ref{figure0}.
The vehicle connected to the Internet transfers the driving data. The server analyzes the data. When the patterns of driving data are not consistent of authorized users, the system detects the theft, and then transfers the alarm message to the vehicle monitoring company and the vehicle owner. The data analysis requires many resources, so it should be conducted in the server-side, as opposed to the data are not carried out in the vehicle.

\section{Methodology and Experiments}

We characterize driver's driving patterns and identify drivers based on the driver's driving characteristics. Fig. \ref{figure1} shows the driver verification framework based on the analysis of driving patterns. The framework consists of four modules that are Data Collection, Feature Preprocessing, Driver Classification Training, and Driver Verification. When the driver starts driving on the road, the data are collected from the sensors in the vehicle. Feature preprocessing module converts the collected data into a new data format to be analyzed in the next module and builds feature vectors that can distinguish drivers. Driver Classification Training module trains the machine learning algorithm using the feature set feed from Feature Preprocessing module. We use Decision Tree, Random Forest, KNN, and MLP, which are shown to have high performance in previous works. The machine learning algorithm detects the unique driving patterns for a driver and builds his or her behavior fingerprints. Driver verification module compares a newcomers' driving patterns and authenticated drivers' patterns and decides whether the driver is authorized or not.

\begin{figure}[b!]
\centering
\vspace{-5mm}
\includegraphics[width=0.93\columnwidth]
{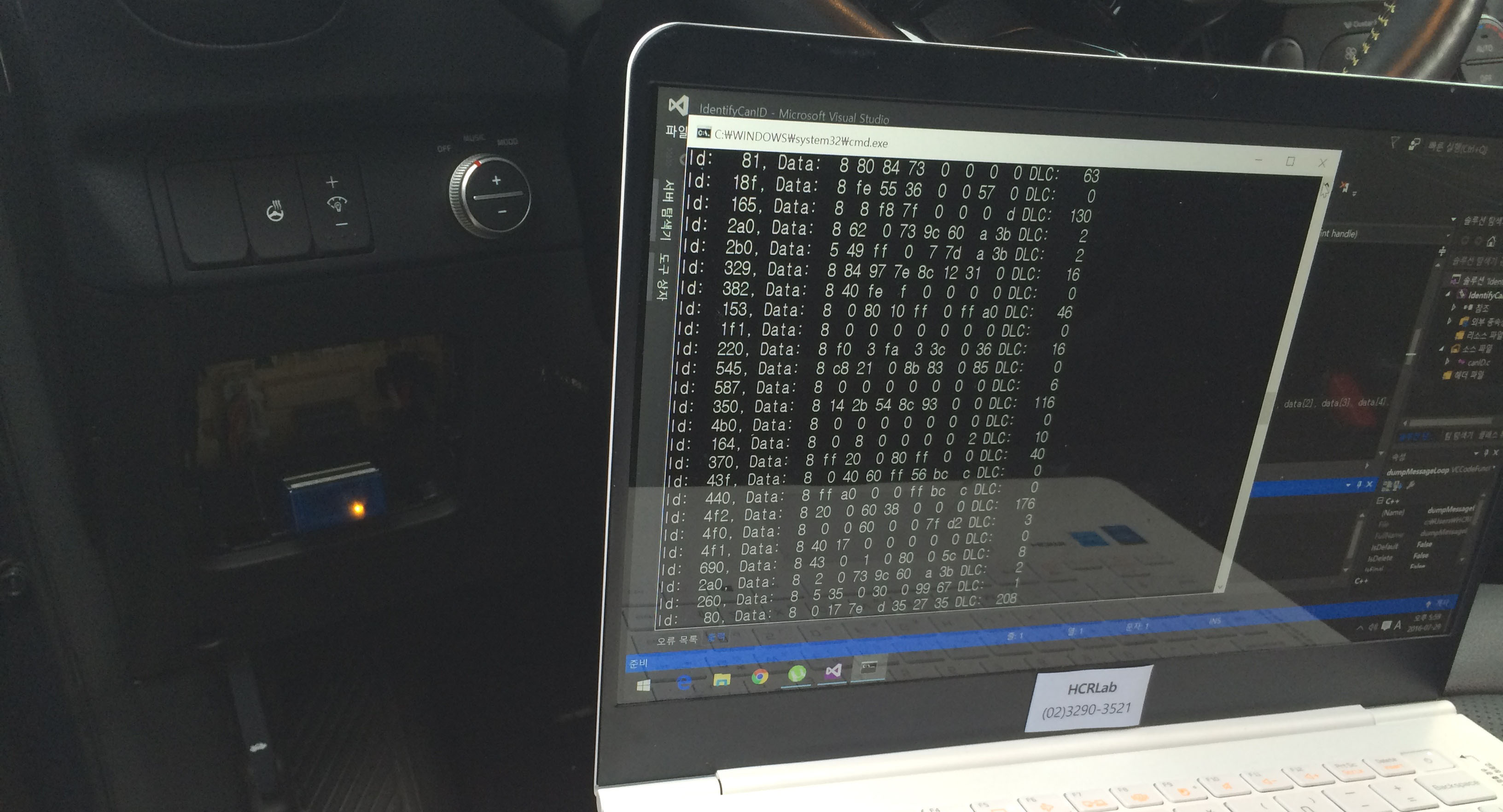}
\vspace{-3mm}
\caption{CAN Data Extraction}
\label{figure10}
\vspace{-3mm}
\end{figure}

\subsection{Data Collection}
\subsubsection{CAN data extraction}
The data are processed from in-vehicle CAN data. This research uses the On Board Diagnostics 2 (OBD-II) and CarbigsP as OBD-II scanner for data collection. 
 Fig. \ref{figure10} is the screenshot of the data extractor in use. 
The recent vehicle has many measurement sensors and control sensors, so the vehicle is managed by Electronic Control Unit (ECU) in it. ECU is the device that controls parts of the vehicle such as Engine, Automatic Transmission, and Anti-lock Braking System (ABS). OBD refers self-diagnostic and reporting capability by monitoring vehicle system in terms of ECU measurement and vehicle failure.
The data are recorded every 1 second during driving. We collected 51 features through the OBD-II.

\subsubsection{Experiment setting}
We extracted the driving data using a recent model of KIA Motors Corporation in South Korea. 
Ten drivers participated in the experiments setting in 4 paths in Seoul.
The driving path consists of three types of city way, motor way and parking space with the total length of 23 km. The experiment is performed since July. 28. 2015. We controlled the time factor by performing experiments in the similar time zone from 8 p.m to 11 p.m on weekdays. 
Ten drivers completed two round trips for reliable classification. 
We collected the data from totally different road conditions. The city way has signal lamps and crosswalks, but the motor way has none. The parking space is required to drive slowly and cautiously. 

\subsubsection{Data set description}
The number of entire features is 51. We derived the min and max value of recorded data per field. We labeled the driving data per driver from ``A'' to ``J''. 
The data that we used has total 94,401 records recorded every second with the size of 16.7Mb in total.
The data set is available at \url{http://ocslab.hksecurity.net/Datasets/driving-dataset}.

\begin{figure}[b!]
\centering
\vspace{-5mm}
\includegraphics[width=0.93\columnwidth]{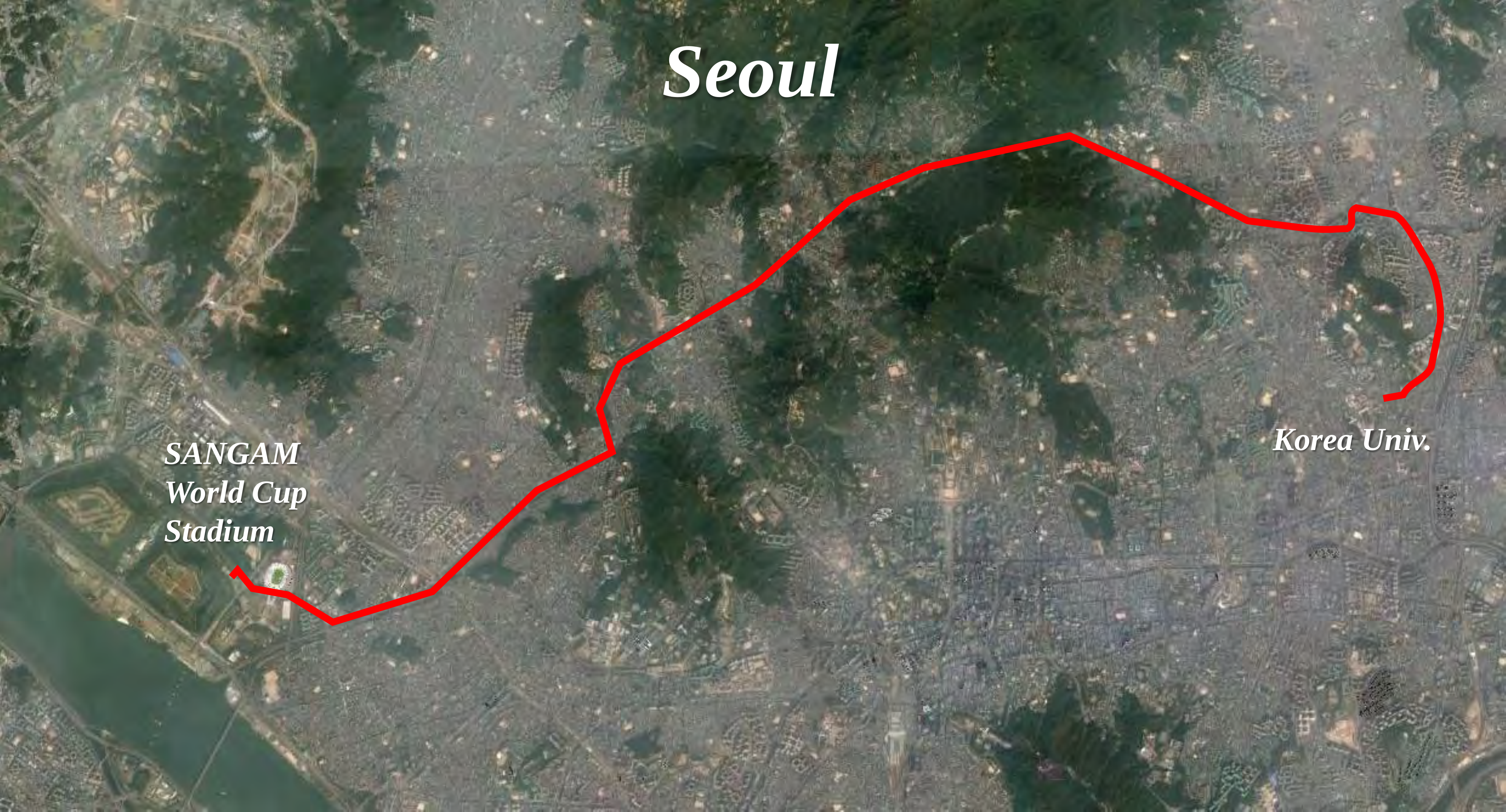}
\vspace{-3mm}
\caption{Drive Loop Location}
\vspace{-3mm}
\label{figure3}
\end{figure}

\begin{table*}[ht!]
\renewcommand{\arraystretch}{1.3}
\vspace{-3mm}
\caption{Features for Driving Pattern Analysis}
\label{table2}
\centering
\begin{tabular}{|m{2.5cm}|m{1.5cm}|c|m{6cm}|c|m{2cm}|}
\hline
\multicolumn{1}{|c|}{Feature}						& Type of vehicle data & \multicolumn{1}{c|}{Range} 		&  \multicolumn{1}{c|}{Description}	&	\multicolumn{1}{c|}{Feature rank} & Previous works \\
\hline
 Long term fuel trim bank1	&	 Fuel	&  -100 -- 100 (\%)	& The correction value being used by the fuel control system in loop modes of operation.	&	 \multicolumn{1}{c|}{1} & \multicolumn{1}{c|}{-}\\
\hline
Intake air pressure			&	Fuel	& 0 -- 255 (kPA)		& A pressure of air inhaled to engine.		&	\multicolumn{1}{c|}{4} & \multicolumn{1}{c|}{-}\\
\hline
 Accelerator Pedal value		&	 Fuel	&  0 -- 100 (\%)		& The degree to which driver is depressing the accelerator pedal.		&	\multicolumn{1}{c|}{9} &  \cite{refer10}, \cite{refer4}, \cite{refer7}, \cite{refer12}, \cite{refer11}\\
\hline
Fuel consumption			&	Fuel	& 0 -- 10000 (mcc)	& The fuel efficiency of an engine.		&	\multicolumn{1}{c|}{11} & \multicolumn{1}{c|}{ \cite{refer9}}\\
\hline
Friction torque				&	Engine	& 0 -- 100 (\%)		& A torque caused by the frictional force that occurs.		&	\multicolumn{1}{c|}{3} & \multicolumn{1}{c|}{-}\\
\hline
Maximum indicated engine torque &  Engine &  0 -- 100 (\%)		&  A calculated value of maximum torque.		&	\multicolumn{1}{c|}{5} & \multicolumn{1}{c|}{-}\\
\hline
Engine torque				& Engine	& 0 -- 100 (\%)		& A force that is spinning engine crankshaft.		&	\multicolumn{1}{c|}{6} & \multicolumn{1}{c|}{ \cite{refer9}}\\
\hline
Calculated load value		& Engine	& 0 -- 100 (\%)		& A percentage of peak available torque. 
		&	\multicolumn{1}{c|}{7} & \multicolumn{1}{c|}{-}\\
\hline
Activation of Air compressor & Engine	& 0 or 1			& A value of air compressor's working.		&	\multicolumn{1}{c|}{8} & \multicolumn{1}{c|}{-}\\
\hline
 Engine coolant temperature	&  Engine	&  -40 -- 215 (\textcelsius)		& The temperature of the engine coolant of an internal combustion engine.	&	\multicolumn{1}{c|}{10} & \multicolumn{1}{c|}{-}\\
\hline
Transmission oil temperature & Transmission & -40 -- 215 (\textcelsius)	& A fluid temperature inside the transmission.		&	\multicolumn{1}{c|}{2} & \multicolumn{1}{c|}{-}\\
\hline
Wheel velocity, front, left-hand &  Transmission &  0 -- 511.75 (km/h) &  The speed of the left front wheel.		&	\multicolumn{1}{c|}{12} & \multicolumn{1}{c|}{-}\\
\hline
Wheel velocity, front, right-hand &  Transmission &  0 -- 511.75 (km/h) &  The speed of the right front wheel.		&	\multicolumn{1}{c|}{14} & \multicolumn{1}{c|}{-}\\
\hline
Wheel velocity, rear, left-hand & Transmission & 0 -- 511.75 (km/h) & The speed of the left rear wheel.		&	\multicolumn{1}{c|}{13} & \multicolumn{1}{c|}{-}\\
\hline
 Torque converter speed		&  Transmission	&  0 -- 16383.75 (rpm)	& A particular kind of fluid coupling that is used to transfer rotating power from a prime mover.		&	\multicolumn{1}{c|}{15} & \multicolumn{1}{c|}{-}\\
\hline
\end{tabular}
\vspace{-5mm}
\end{table*}

\subsection{Feature Preprocessing}

Feature Preprocessing transforms collected data into new information that can be used in analysis and classification algorithm. 

\subsubsection{Feature selection}
We excluded identical and extraneous features. For example, Engine torque value is identical to correction of Engine torque value. After deleting redundant features, we performed feature selection to exclude highly correlated features for improving the performance in terms of the accuracy and speed. 
We selected 15 features from 51 features. 
Table \ref{table2} shows the features selected from feature selection. 
We reported the rank representing the importance of features derived using InfoGainAttributeEval evaluation method, one of Ranker search methods among the feature selection methods implemented in Weka \cite{hall2009weka}. 


\subsubsection{Feature normalization}

The features have different scales, so they are normalized to be equally treated in the classification algorithm. Especially, the normalization process is necessary for the algorithm run based on the distance between data, e.g. KNN algorithm. 

\begin{equation}
X_i = \frac{(x_i-min(x_i))}{max(x_i) - min(x_i)}
\end{equation}

We use Equation (1) to normalize the feature to be ranged between 0 and 1. In the equation, min is the minimum value of feature data, and max is the maximum value of the feature respectively.

\subsubsection{Statistical features}
\begin{figure*}[ht!]
\centering
\vspace{-3mm}
\includegraphics[width=\textwidth]{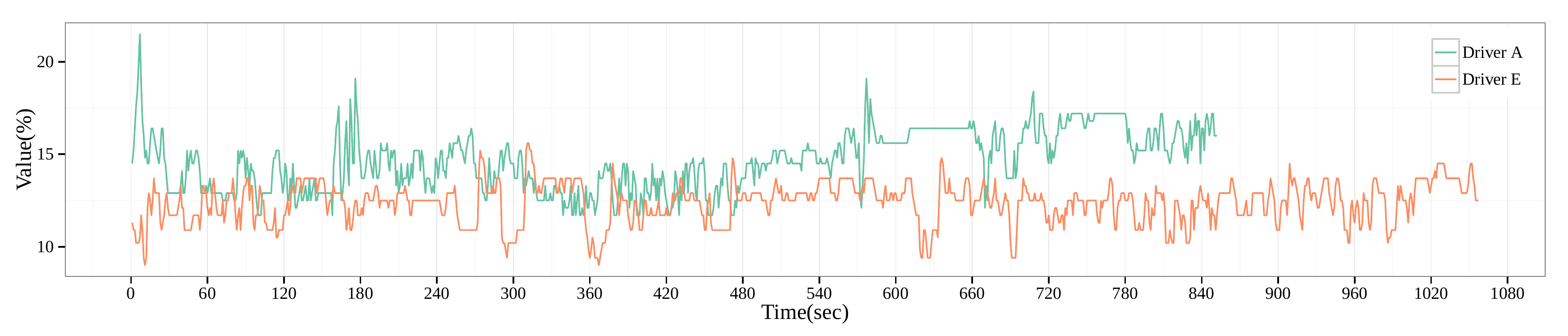}
\vspace{-7mm}
\caption{Time series of extracted CAN data}
\vspace{-5mm}
\label{timeseries}
\end{figure*}

Fig. \ref{timeseries} shows time--series patterns of CAN data per a user. In a real driving condition, CAN data naturally have fluctuation.

\begin{figure}[ht!]
\centering
\vspace{-3mm}
\subfigure[Long Term Trim Bank1]{\includegraphics[width=0.49\columnwidth]{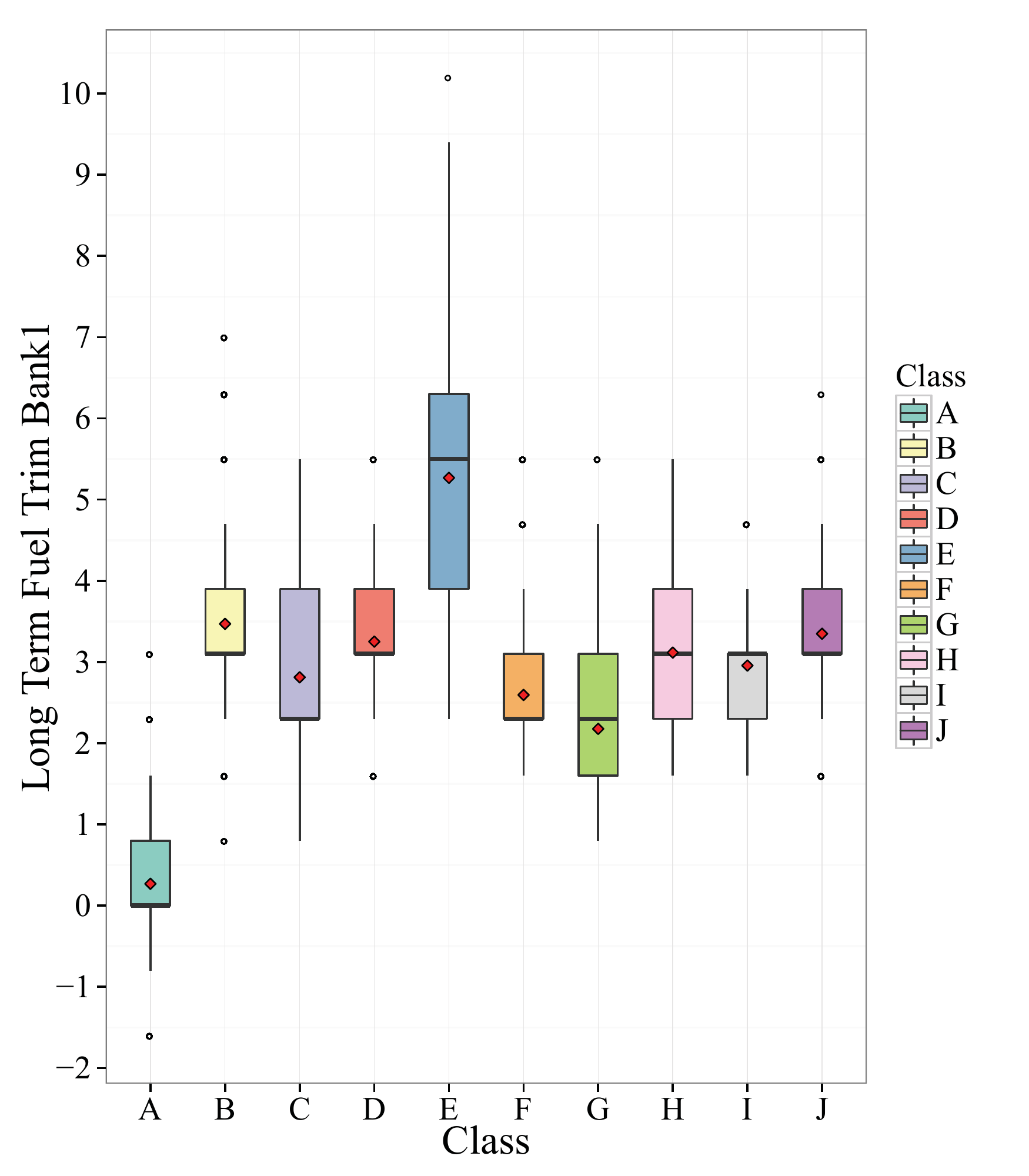}}\hfill
\subfigure[Transmission Oil Temperature]{\includegraphics[width=0.49\columnwidth]{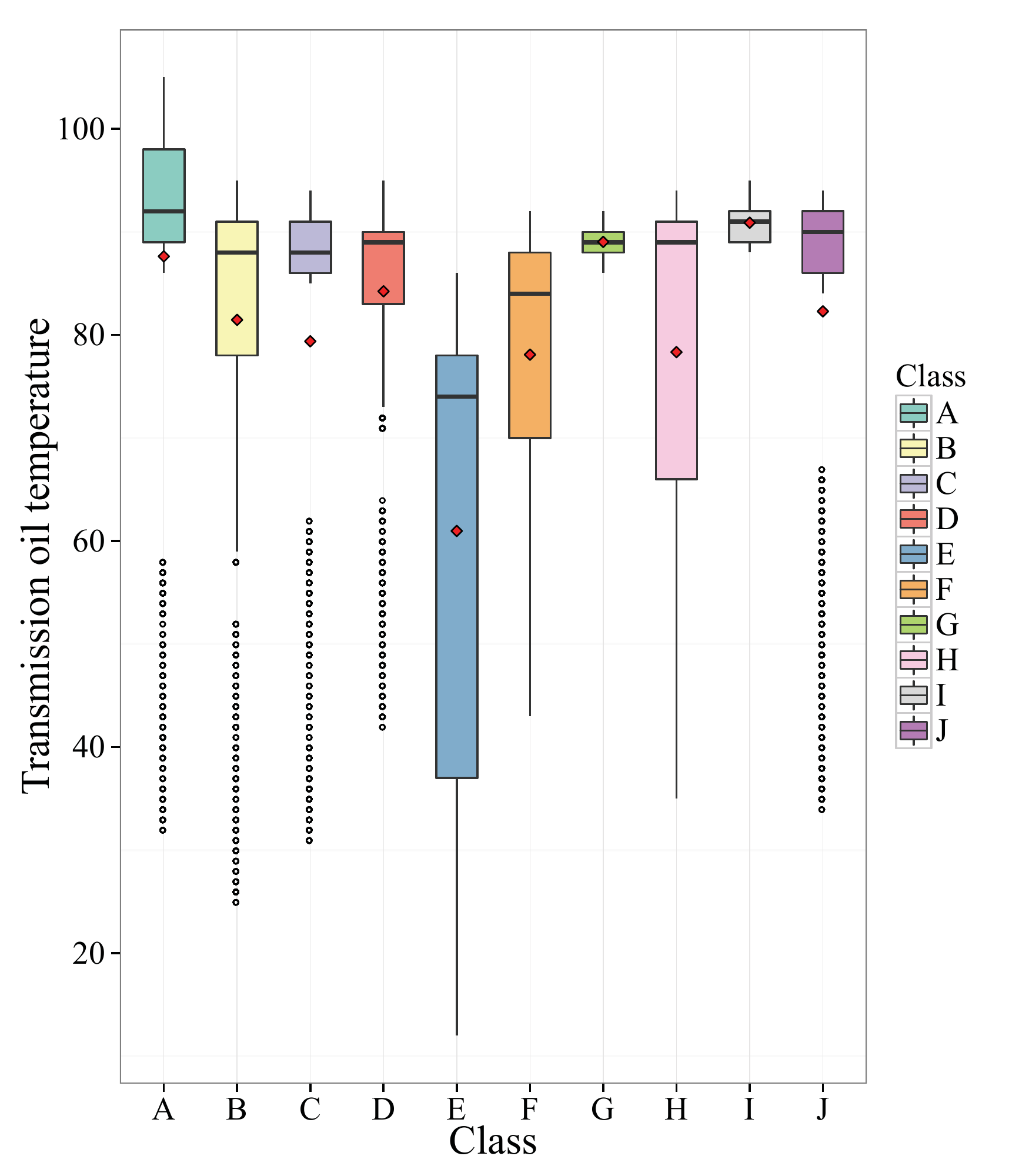}}\hfill
\vspace{-2mm}
\caption{Feature Distribution according to drivers in city road}
\label{figure9}

\centering
\subfigure[Long Term Trim Bank1]{\includegraphics[width=0.49\columnwidth]{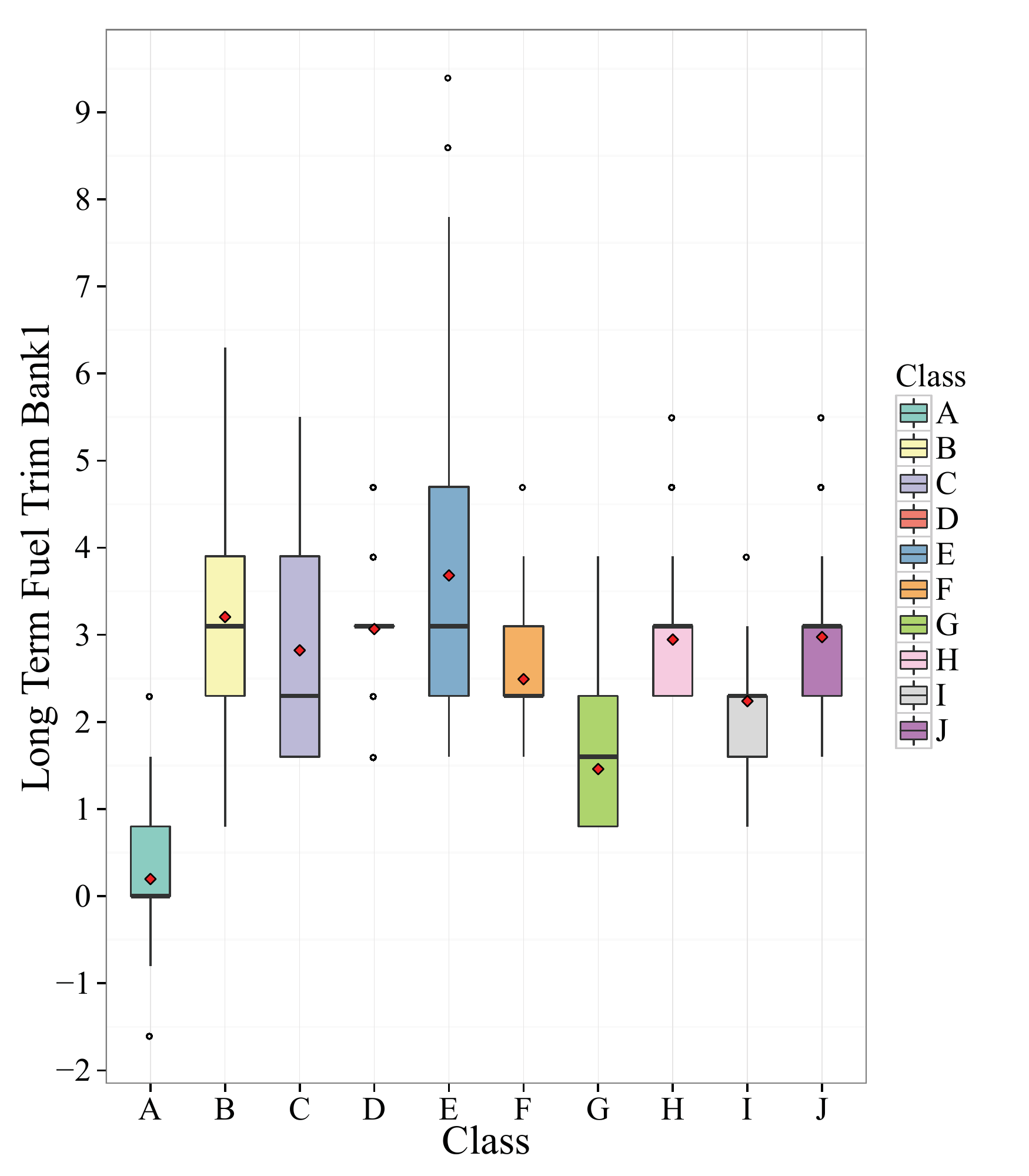}}\hfill
\subfigure[Transmission Oil Temperature]{\includegraphics[width=0.49\columnwidth]{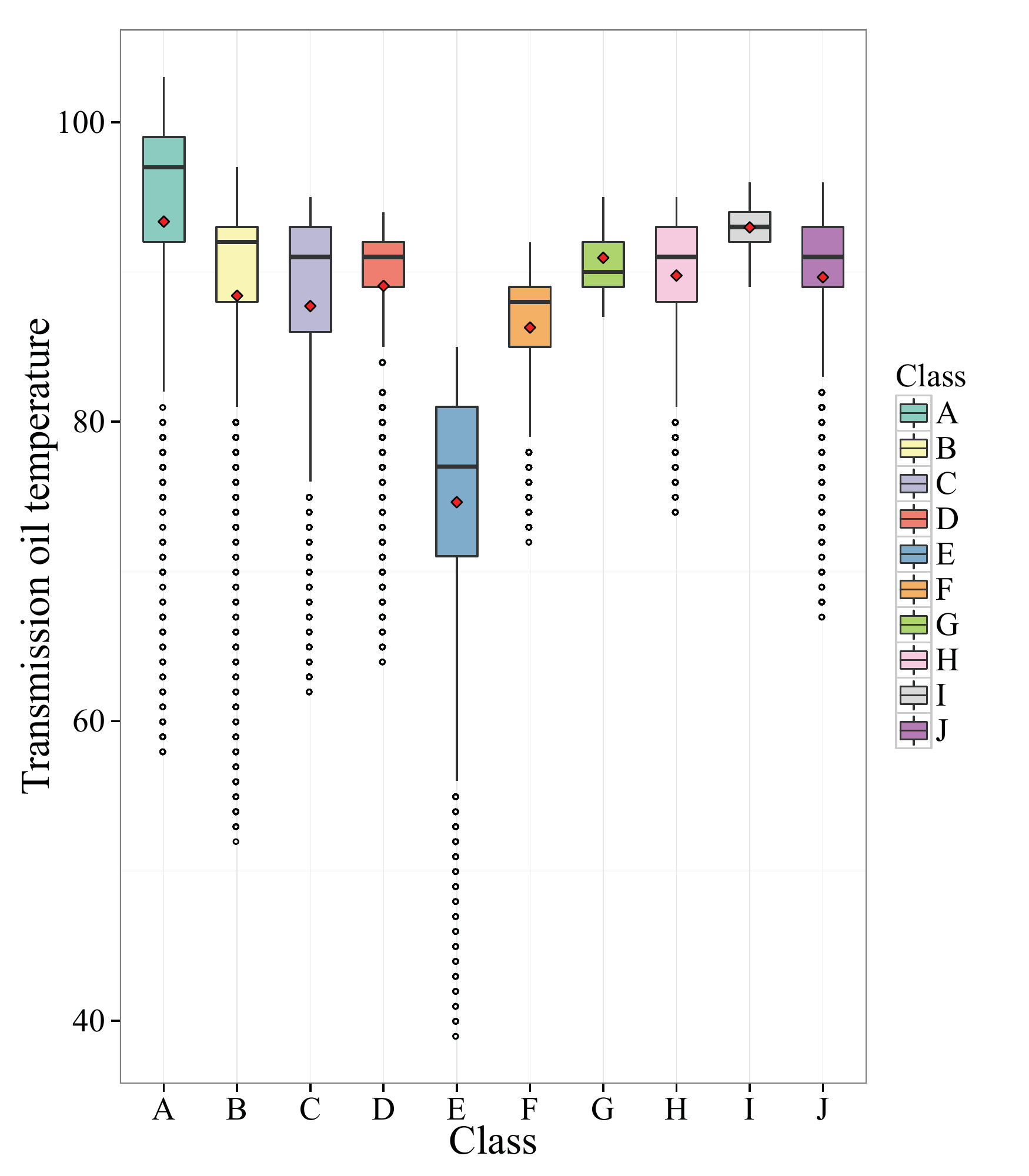}}\hfill
\vspace{-2mm}
\caption{Feature Distribution according to drivers in vehicle road}
\label{figure11}

\centering
\subfigure[Long Term Trim Bank1]{\includegraphics[width=0.49\columnwidth]{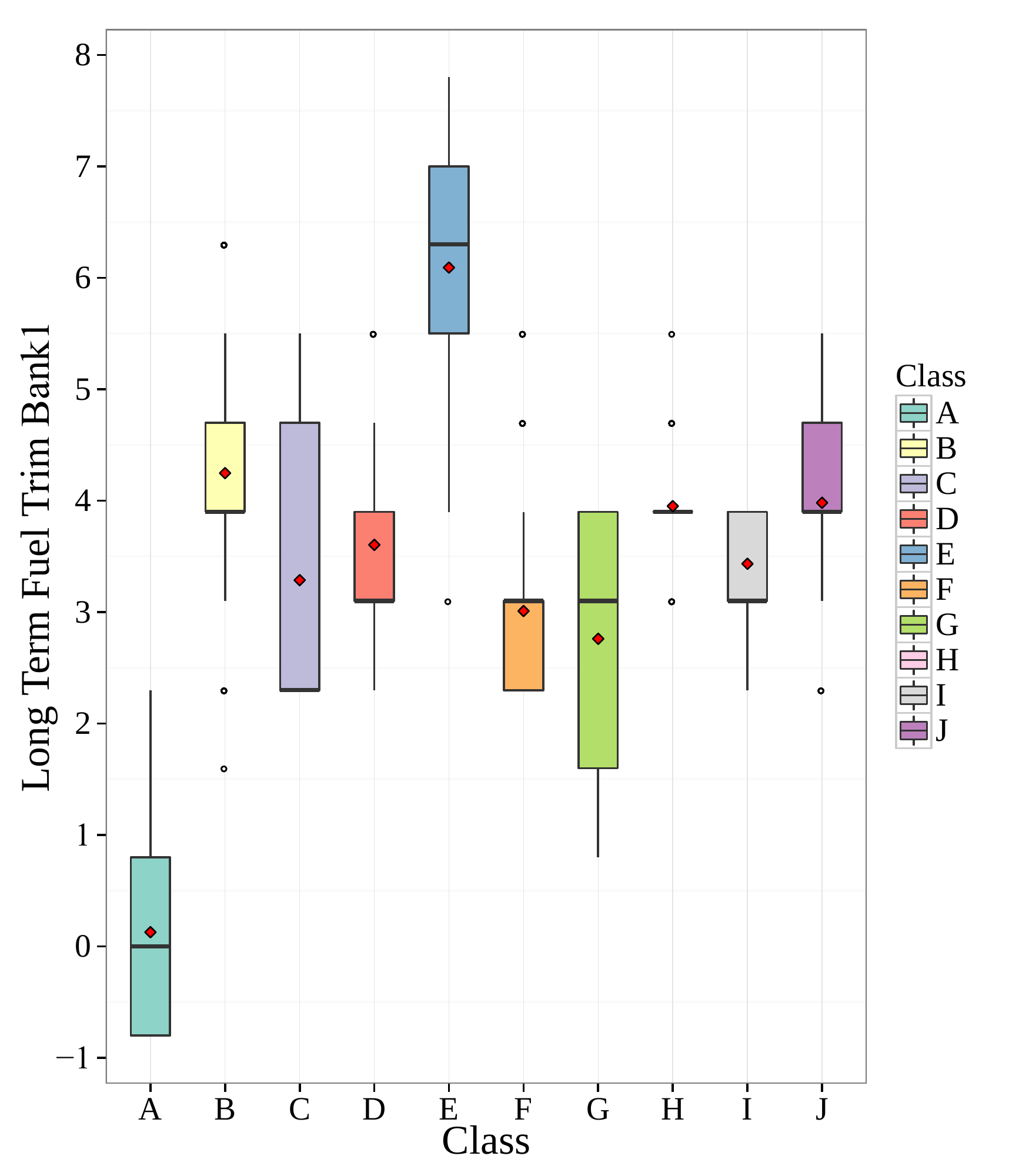}}\hfill
\subfigure[Transmission Oil Temperature]{\includegraphics[width=0.49\columnwidth]{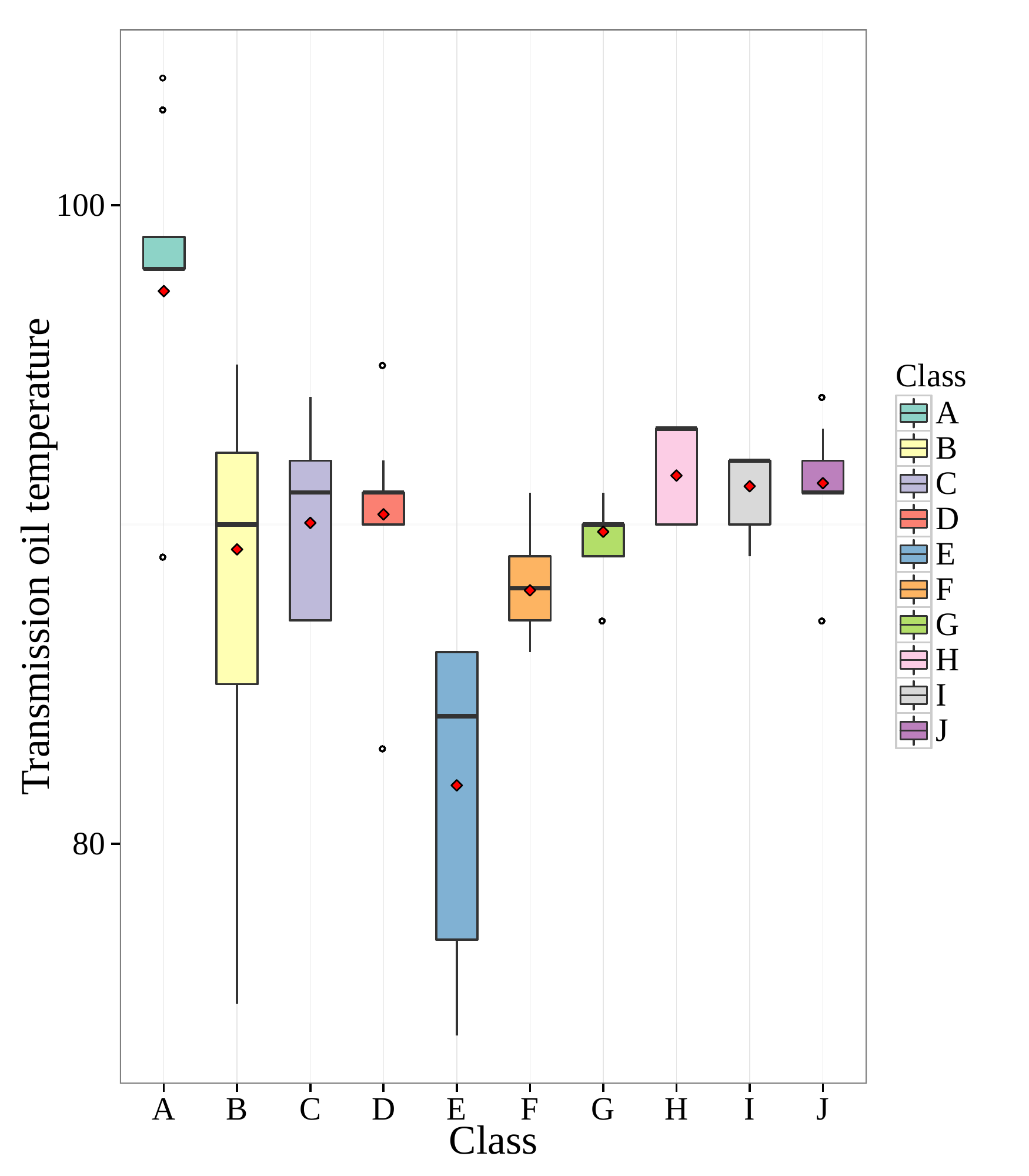}}\hfill
\vspace{-2mm}
\caption{Feature Distribution according to drivers in parking lot}
\label{figure12}
\vspace{-4mm}
\end{figure}

We then checked the distribution of the features according to the road types and displayed the distributions of two major features in Fig. \ref{figure9}, Fig. \ref{figure11}, and Fig. \ref{figure12}.
In Fig. \ref{figure9}-a, `Long term fuel trim bank1', which is the correction value of `Short-term fuel trim bank1', checks the condition of engine fuel.
Fuel trims are the percentage of change in fuel over time. This value is changing according to driving environment such as start-up, idling in heavy traffic, cruising down the highway, etc \cite{obd_fuel_trims} and driver's driving pattern in such conditions. 

Most values of `Long term fuel trim bank1' range between -10\% and 10\%. In Fig. \ref{figure9}-a and b, drivers have different feature distributions from other drivers. Furthermore, we find that feature distribution can be characterized by the mean and standard deviation per a driver. 

Among many features, `Long Term Fuel Trim Bank1' and `Transmission oil temperature' are the most visually differentiated features to classify drivers.

\begin{figure}[b!]
\centering
\vspace{-6mm}
\includegraphics[width=7.6cm]{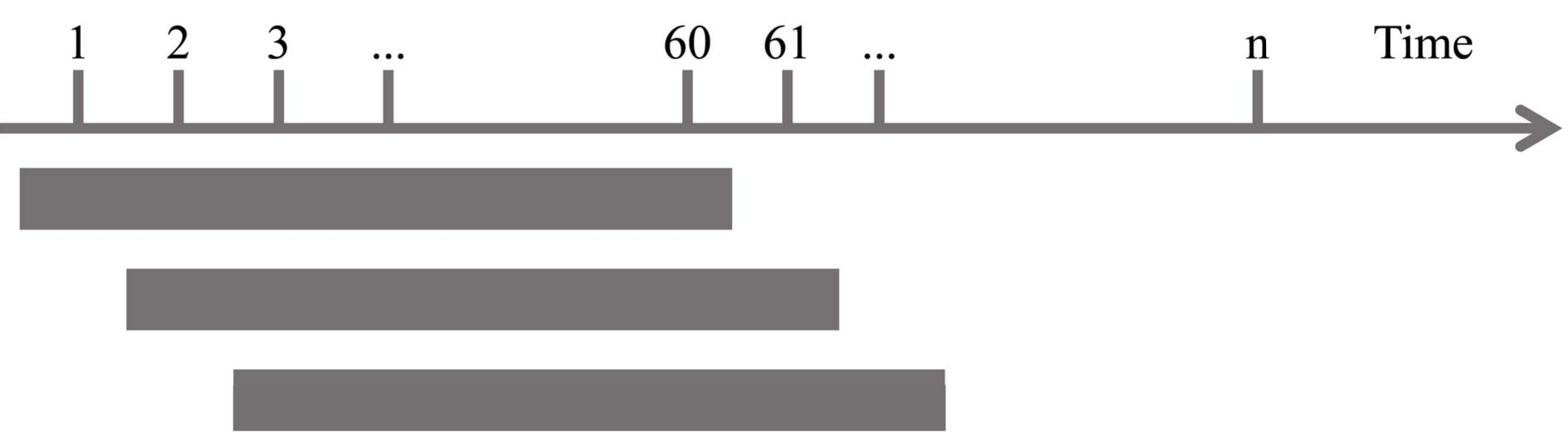}
\vspace{-2mm}
\caption{Time Segmentation for Statistical Feature Extraction}
\vspace{-4mm}
\label{figure2}
\end{figure}

To minimize the effect of fluctuation of feature values and characterize the feature distribution effectively, we adopt the statistical values including mean, median, and standard deviation. 
We also adopt the median for highly skewed values.
The statistical features are derived every period determined by sliding window. Fig. \ref{figure2} shows how to set the sliding window. If the sliding window size is set 60, statistical features are firstly calculated from 1 to 60 seconds. The second period is set from 2 seconds to 61 seconds.
We derive three statistical features for an original feature and complete the feature set with the original features and their statistical features. Thus, a record has 15 original features and 45 statistical features in total.

\subsection{Driver Identification}

\subsubsection{Algorithm selection}
In the driver identification phase, we consider the algorithms that show good performance in previous works, e.g. the decision tree, random forest, k-nearest neighbors algorithm (KNN), and multi--layer perceptron (MLP). The decision tree is the classification algorithm that the data recursively is divided into small parts in terms of attributes of the feature that guarantees the highest information gain. The decision tree has the pros and cons. It is easy to interpret, so it is intuitively appealing. However, it has relatively low accuracy than other classification algorithms. Random forest is an ensemble learning method consisting of many decision trees. 
Random forest corrects the over-fitting problem of the decision tree. Random forest has more complexity than the Decision Tree, but they have high accuracy than the decision Tree. 
KNN is a classification algorithm that classifies the data based on the distance between training examples. It has no assumptions about data, and it guarantees the high accuracy, but it is computationally expensive.
MLP is a classification algorithm generating hyper-variables in a way that it fully connects the nodes in lower-layer to nodes in the higher-layer. MLP has the high accuracy than other algorithms, but it is computationally expensive. Also, it can cause an over-fitting and under-fitting problem.

The classification algorithms generate the driver identification model. 

\subsubsection{Identification results}
We performed the driver identification training every second since driving patterns are recorded every second. 
We adopt 10-fold cross-validation to have not only high accurate, but also high generalization ability. It divides the data into 10 parts, trains the model with 9 parts and evaluates the model with one remaining part. When a new driving data is fed, the evaluation module classifies it into one of the pre-defined classes with the highest similarity.
We adopted four algorithms, the decision tree, random forest, KNN, and MLP implemented in WEKA.
Table \ref{table3} shows the classification accuracy in 3 road types. All algorithms have accuracy over 99\%. The vehicle way has the higher accuracy than the city way and parking lot.

\begin{table}[ht!]
\renewcommand{\arraystretch}{1.3}
\vspace{-2mm}
\caption{Average Accuracy Of Driver Identification}
\label{table3}
\vspace{-1mm}
\centering
\begin{tabular}{|c|m{1.3cm}|m{1.3cm}|m{1.3cm}|m{1.3cm}|}
\hline
Road type	&	Decision Tree		&	KNN	&	Random Forest	&	Multilayer perceptron	\\
\hline
City way	&	0.987	&	0.963	&	0.998	&	0.948	 \\
\hline
Vehicle way	&	0.990	&	0.984	&	0.998	&	0.989	\\
\hline
Parking lot	&	0.978	&	0.925	&	0.993	&	0.956	\\
\hline
\end{tabular}
\vspace{-2mm}
\end{table}

\subsection{Driver Verification}

\begin{figure}[hb!]
\centering
\vspace{-7mm}
\includegraphics[width=0.93\columnwidth]{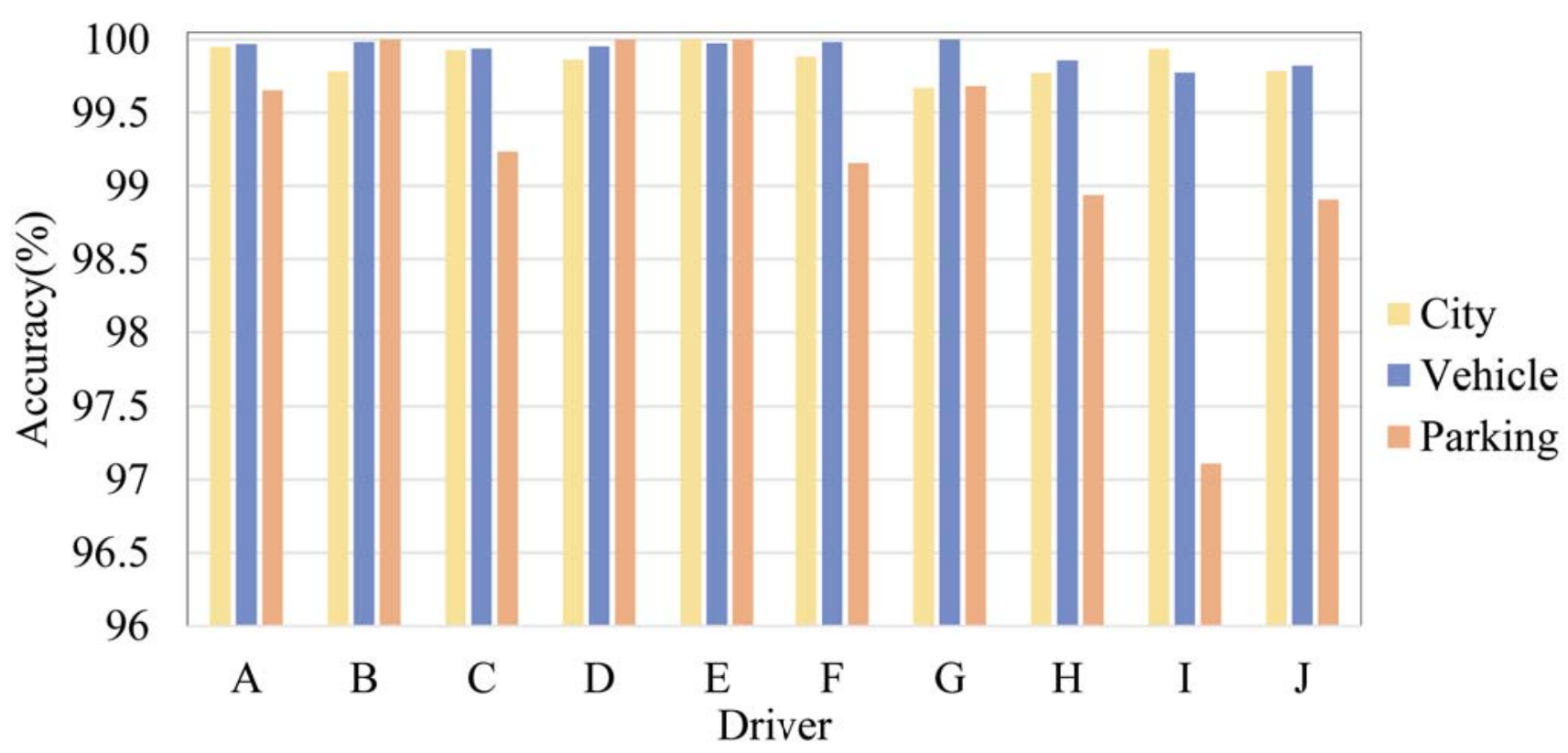}
\vspace{-3mm}
\caption{Result of Random Forest in Route type}
\vspace{-4mm}
\label{figure13}
\end{figure}

The authorized driver will be recognized by classification algorithms, and the unauthorized driver will not.
Using the classification model trained from authorized users, we test whether the user is classified into the pre-defined classes, e.g. authorized drivers, or not. The testing process is performed based on the similarity between a new data samples and pre-defined classes. Because we test users every second, we can derive the classification accuracy for a time window. 
We need to set up the threshold value of the similarity for issuing an alert. The threshold value is the minimum accuracy generated by the algorithms trained on authorized drivers. 
If the similarity measured by all algorithms exceed the threshold value, the driver is verified as the authorized driver. Otherwise, the server connected to the vehicle sends a warning message to the owner or the vehicle control center to protect robbery.
As shown in Fig.\ref{figure13}, for each driver, we tracked the detection accuracy. Overall, the detection in the parking lot is lower than those in other road types. The threshold value to issue an alert should be set to 0.97.

\section{Detection Sensitivity Test}

\subsection{Real-time detection}

\begin{figure}[t!]
\vspace{-4mm}
\centering
\subfigure[City way]{\includegraphics[width=0.83\columnwidth]{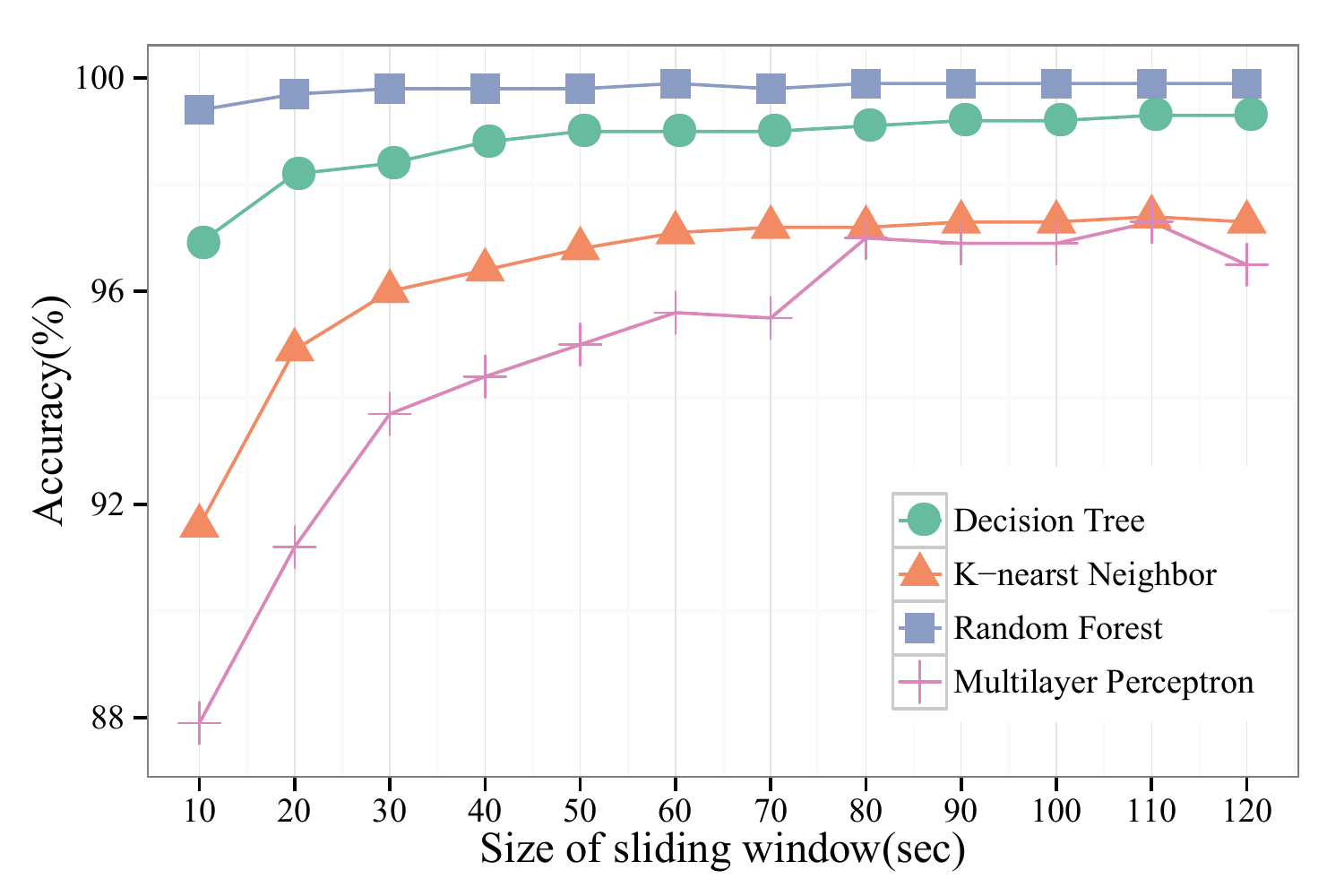}}\hfill
\subfigure[Vehicle way]{\includegraphics[width=0.83\columnwidth]{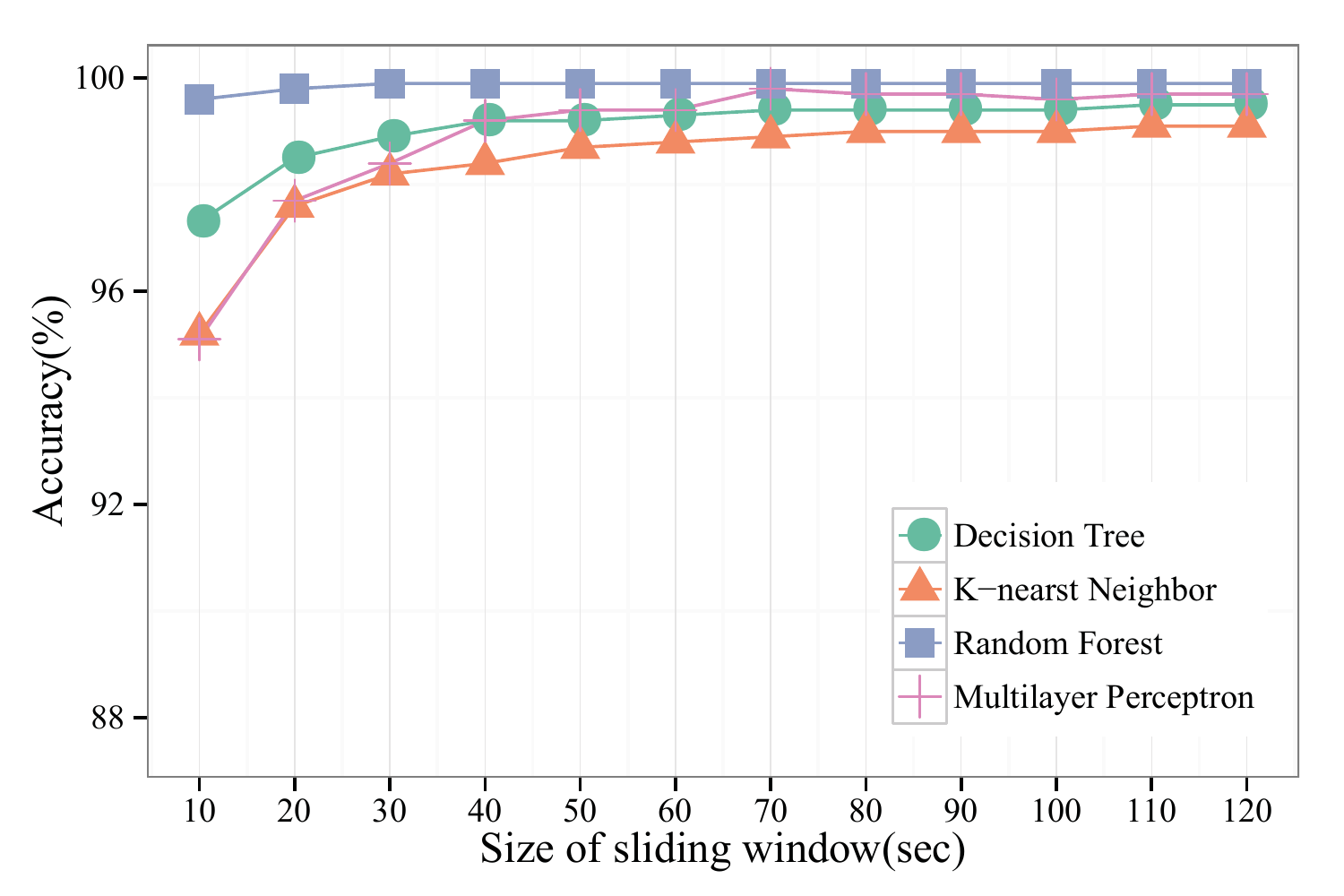}}
\subfigure[Parking lot]{\includegraphics[width=0.83\columnwidth]{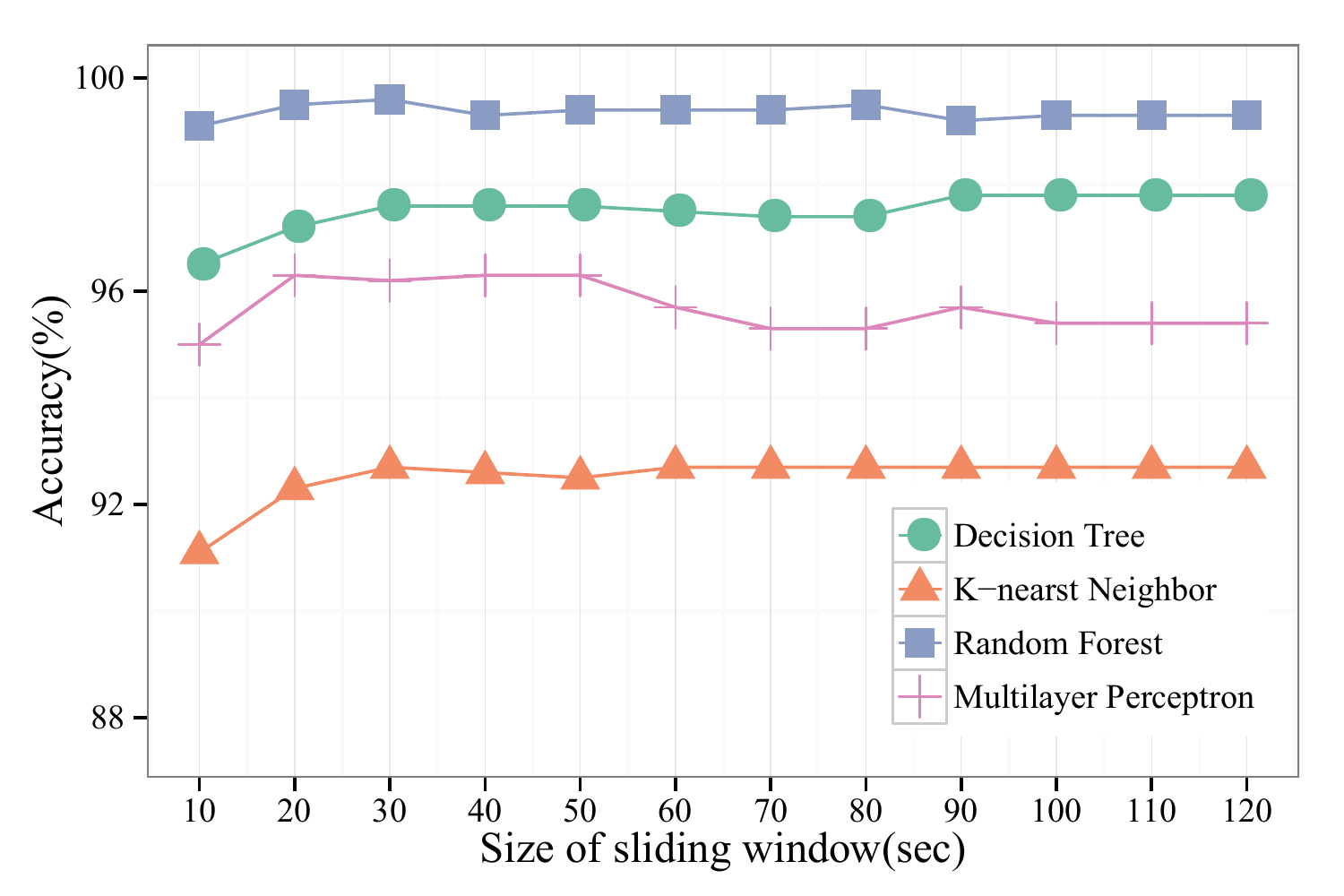}}\hfill
\caption{Algorithm Accuracy with Sliding Window Changes according to the way}
\vspace{-6mm}
\label{figure4}

\end{figure}

For fast detection and notification to the user in case of auto-theft, we need a real-time detection system. 
We tested our system varying the size of data feed to driver identification training. 
For fast detection, the window size should be short, but for reliable detection, it should be long. We checked when the detection performance becomes reliable according to the road types and the detection algorithm types as shown in Fig. \ref{figure4}. 
This time-sensitivity test can also be used to find out an optimal value of accuracy that compensates above two conditions. 
We notice that Multilayer perceptron and KNN algorithm are not reliable in the initial phase of detection. As time goes on, the performance becomes reliable. 
The similarity-based method requires enough data to identify drivers, but the algorithm that operates by determining threshold values of features according to drivers exhibits stable performance with only initial data. This indicates that we need a sophisticated machine learning algorithm to detect auto-theft rather than similarity-based method. 
We also notice that the difference of algorithm accuracy with sliding window changes. Accuracy is increased with the rise of sliding window size in city way and vehicle way. However, in parking lot, accuracies are similar regardless of the changing size of sliding window. The most of the vehicle way accuracies have the higher than city way or parking lot accuracies. Moreover, there is a difference in accuracy according to the algorithms. The performance of random forest is better than other algorithm's them in all the type of ways. In comparison, KNN has the lower in vehicle and city way. 

If we set the sliding window size 60, we can guarantee the high accuracy of the algorithms in relatively short time. If we want to get the higher accuracy, we need to set up 120 of sliding window size. However, it takes longer to issue alert to owners.

\subsection{Feature dependency}

We compare the accuracy our feature set with the feature set in previous works. We also compare the feature set including original features and the feature set including statistical features. 
Features used in previous works consists of the accelerator pedal, braking pedal, steering wheel, vehicle speed, engine speed, gear, throttle position, and engine coolant temperature. Table \ref{table5} shows the comparison of accuracy with different feature sets. 
Our feature set shows much higher accuracy than the feature set used in previous works. 
We can significantly improve performance when we add statistical features.
We diversify the feature set and increase the accuracy through feature processing of statistical features generated for an appropriate size of sliding window.

\begin{table}[ht!]
\renewcommand{\arraystretch}{1.3}
\caption{Average accuracy comparison according to the changing feature set}
\label{table5}
\centering
\begin{tabular}{|m{0.2cm}|m{1.9cm}|m{1cm}|m{1cm}|m{1cm}|m{0.9cm}|}
\hline
\# & Feature set	&	Decision Tree	&	KNN	&	Random Forest	&	MLP\\
\hline
1&Our research's feature set		&	0.938		&	0.844		&	0.961	&	0.747\\
\hline
2&Our research's feature set + statistical feature		&	0.984		&	0.957		&	0.996	&	0.964\\
\hline
3&Other research's feature set	&	0.459		&	0.422		&	0.528	&	0.302\\
\hline
4&Other research's feature set + statistical feature		&	0.852			&	0.790			&	0.938		&	0.641\\
\hline
\end{tabular}
\end{table}

\section{Conclusion}

We proposed an anti-theft method based on driving pattern of the vehicle. Our experiment results show that drivers have own characteristics in driving. 
Our system extracts the mechanical feature from automotive parts in the vehicle, selects important features, and extracts statistical features after optimizing the window size.
Our experiments showed that in-vehicle network data including `Long term fuel trim bank1', `Transmission oil temperature' are important in identifying drivers.
These features reflecting driving patterns are difficult to detour because the driver can not manipulate these data.

 We found the optimal size of sliding window. The size of sliding window plays a major role in reducing the data resource and increasing accuracy. The size of sliding window is set 60 to trade off time cost against accuracy. This indicates that it will take 60 seconds until the owner receives the alert on break-in.
For future research, first of all, we will make more people to join in experiments to improve generality of our model. 
Second, our research can be utilized not only the driver authentication but also anomaly detection. 
We are planning to perform the experiment after checking the driver's fatigue and extend our research in detecting abnormal behaviors such as sleepiness and drinking. We will examine the distribution of features and set control limit of feature per use. Then, we will correlate the feature distribution and the abnormal behaviors.
Thirdly, we will try mimicry attack that a driver tries to mimic the other driver's driving pattern and will see our method can be defeated or not. 



\ifCLASSOPTIONcompsoc

  \section*{Acknowledgments}
  
\else

  \section*{Acknowledgment}
\fi

This work was supported by Samsung Electronics, Co., Ltd.

\bibliographystyle{IEEEtran}
\bibliography{references}

\end{document}